\documentclass[aps,prl,twocolumn,groupedaddress]{revtex4}
\usepackage{graphicx}

\begin{document}

\title{Dynamics of text generation with realistic Zipf distribution}

\author{Dami\'an H. Zanette}
\email{zanette@cab.cnea.gov.ar} \affiliation{Consejo Nacional de
Investigaciones Cient\'{\i}ficas  y T\'ecnicas, Centro At\'omico
Bariloche and Instituto Balseiro, 8400 San Carlos de Bariloche,
R\'{\i}o Negro, Argentina. }

\author{Marcelo A. Montemurro}
\email{mmontemu@ictp.trieste.it} \affiliation{Abdus Salam
International Centre for Theoretical Physics, Strada Costiera 11,
34014 Trieste, Italy}

\date{\today}

\begin{abstract}

We investigate the origin of Zipf's law for words in written
texts by means of a stochastic dynamical model for text
generation. The model incorporates both features related to the
general structure of languages and memory effects inherent to the
production of long coherent messages in the communication
process. It is shown that the multiplicative dynamics of our model
leads to rank-frequency distributions in quantitative agreement
with empirical data. Our results give support to the linguistic
relevance of Zipf's law in human language.

\end{abstract}

\pacs{05.40.-a, 89.75.Fb, 89.75.Da}

\maketitle

Natural languages are complex systems that have evolved as
effective dynamical structures, capable of codifying and
transmitting highly nontrivial information. The foremost example
of the kind is the genetic code, which evolved as a very
efficient way of coding instructions to replicate living things
\cite{smith}. Another example of a natural code is human
language. Although it evolved in shorter time scales than the
genetic code and was subject to a noisier environment, it also
achieved its present complexity by the natural process of
evolution \cite{nowak}. In recent years we have witnessed a
gradual extension of the ideas and methods of statistical physics
onto a vast range of complex phenomena outside the traditional
boundaries of physical science. When language is studied with
tools originally developed within statistical mechanics, a very
rich structure emerges at levels ranging from simple word counts
to large scale organizational patterns that span many thousand of
words \cite{vocabulary,EN,kanter,mmdz,mmpp}. The use of statistics
applied to human language records draws a picture of its
macroscopic structure, at a level of description that may disclose
traces of its evolutionary history and provide information on the
complex processes behind its generation by the brain.

One the most generic statistical properties of written language is
established by Zipf's law, in the form of a mathematical relation
between the rank of each word in a list of all the words used in
a text ordered by decreasing frequency, and its frequency
\cite{zipf1,zipf2}. Defining the normalized frequency of a word
in a text as $f(r) = n(r)/T$, where $n(r)$ is  the number of
occurrences of the word at rank $r$ in the frequency-ordered list
and $T$ is the total number of words in the text, Zipf's law reads
\begin{equation}
\label{zipf} f(r)\propto r^{-z}.
\end{equation}
In the original formulation of this empirical law, the exponent
$z$ was taken to be exactly equal to one. If, instead, the
exponent is assumed as a parameter and fitted to empirical data
it may take on values different from unity.

After more than fifty years since the discovery of Zipf's law in
human language, its ultimate origin remains elusive. However, the
remarkable ubiquity of this law over diverse languages suggests
that its origin must be looked for in very general probabilistic
aspects associated with the process of language generation. Two
models explaining Zipf's law are worth mentioning, which
represent two very different positions with respect to the
linguistic significance of the law. The first one, a stochastic
process put forward by H.~Simon \cite{simon}, simulates the
dynamics of text generation as a multiplicative process that
leads to Zipf's law for asymptotically long texts. The second
model is due to B.~Mandelbrot \cite{mandel}, who explained Zipf's
law as a static feature of the statistical structure of random
symbolic sequences. While Simon's model gives Zipf's law a
nontrivial linguistic significance, Mandelbrot's explanation
renders the law as a mere property of a random array of symbols.
Let us also recall that, recently, it has been proposed that
Zipf's law could  be the result of Markov processes underlying
text-generation dynamics \cite{kanter}. However, the analytical
results were only tested against the text of the Bible, which is
in fact a large collection of relatively short writings and whose
statistics would therefore be subject to strong random
fluctuations.

In this paper, we present a dynamical model that explains the
empirical behavior of words frequencies as a consequence of two
processes acting at different scales: a global memory effect
driven by context, essentially related to the interplay between
multiplicative and additive dynamics in word selection, and a
local grammar-dependent effect associated with the appearance of
word inflections. This model is able to reproduce realistic Zipf
distributions and, additionally, has a linguistic interpretation.
We show that the controversy between Simon's and Mandelbrot's
views may be solved after explaining the particular deviations
that empirical distributions exhibit with respect to the original
form of Zipf's law.

\begin{figure}
\begin{center}
\includegraphics[width=\columnwidth,keepaspectratio=true]{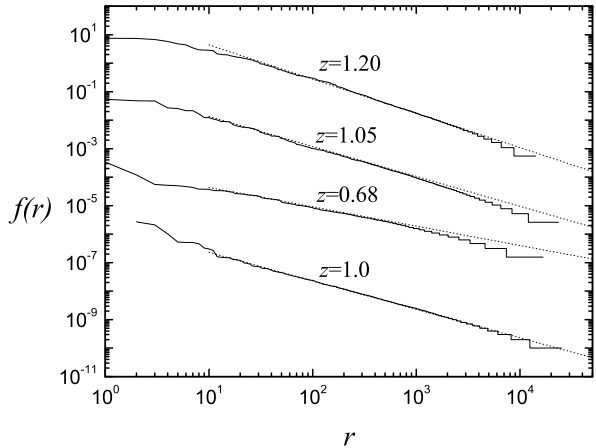}
\caption{From top to bottom, rank-frequency distributions from
David Copperfield by Charles Dickens, Don Quijote de la Mancha by
Miguel  de  Cervantes, and {\AE}neid  by  Virgil, respectively.
The fourth distribution represents a realization of   Simon's
model. The dotted lines stand for pure power laws, $r^{-z}$, fitting
the distributions in the intermediate interval of ranks
($20<r<10^3$). For clarity, the curves are displaced in the
vertical scale.} \label{fig1}
\end{center}
\end{figure}

As advanced above, empirical rank-frequency distributions from
real text sources are more complex than the original form of
Zipf's law, Eq.~(\ref{zipf}).  The three uppermost curves in
Fig.~\ref{fig1} show, from top to down, the rank-frequency
distribution of words for three literary works written in English,
Spanish and Latin, respectively. In all three cases the
qualitative features of the distributions are similar, but some
quantitative differences are worth mentioning. Above a short range
corresponding to the most frequent words, where the behavior of
$f(r)$ is rather irregular, the distributions decay as a power law
in an interval of ranks spanning, for these texts, little less
than two decades. Immediately following the power-law regime, the
distributions fall slightly faster, as can be seen by comparing
with the pure power laws shown in the graph as dot lines. Note
that the values of the exponent $z$ measured in the power-law
regime deviate appreciably from unity in the English and Latin
texts. This phenomenology seems to be generic, since consistent
quantitative results are found for other works in the same
languages.

Simon's model \cite{simon} intends to capture the essential
features of actual text generation by specifying how words are
added to the text, as follows. Suppose that at each step $t$ a
new word is added to the text beginning with just one word at $t
= 1$ such that, at any step, the length of the text is $t$. With a
fixed probability $\alpha$, a new word not present in the text is
added at $t+1$ or, with the complementary probability, the new
word is chosen among the previous $t$ words at random. Since the
probability of repeating words that have already appeared is
taken to be proportional to their number of previous occurrences,
this process establishes a strong competition among different
words. It can be shown that the long time rank-frequency
distribution arising from Simon's model is a power law of the form
of Eq.(\ref{zipf}) with $z=1-\alpha$. We have included in
Fig.~\ref{fig1} the results of a simulation of Simon's model,
with $\alpha=0.01$. Comparison with distributions obtained from
real texts reveals qualitative agreement, though quantitative
differences are apparent also. In particular, Simon's model does
not reproduce the faster decay at low frequencies, and is unable
to explain power-law exponents larger than unity, as observed in
English and Spanish texts. We show in the following that the model
admits to be modified on the base of linguistically sensible
assumptions, in such a way as to correct such differences.

In  Simon's model, new words are introduced at a constant rate
$\alpha$, such that the vocabulary size at step $t$ is, on the
average, $V_t=\alpha t$. Empirical data show however that the
growth of vocabulary in real texts is typically sublinear
\cite{vocabulary}, and may be approximated by
\begin{equation}
\label{growth} V_t=\alpha t^{\nu},
\end{equation}
with $0<\nu<1$. This generalized form of $V_t$ correspond to a
rate of introduction of new words given by $\alpha \nu
t^{\nu-1}$. This provides the first modification to Simon's
model, introducing a new parameter $\nu$. Instead of using a
constant probability for the introduction of new words, we take a
step--dependent probability $\alpha_0 t^{\nu-1}$, with
$\alpha_0=\alpha \nu$. We distinguish two main factors that
affect the value of $\nu$. One factor depends on author and style
and may explain small differences of vocabulary growth between
works in the same language. However, a stronger effect on $\nu$
results from the degree of inflection of each language: highly
inflected languages like Latin---where a single root produces,
through declination and conjugation, many different
words---require a higher value of $\nu$ than poorly inflected
languages, such as English. In a text written in a highly
inflected language, word forms will proliferate significantly
faster than in languages with few inflected forms.

The original dynamics of Simon's model implies that, when a word
has  to be taken from the text written so far, the probability
that a word is repeated is proportional to $n_i$, the number of
its previous occurrences. However, newly introduced words do not
have a clearly defined influence on the context yet, and the
probability that they are used again should be treated in a
slightly different way. Our second modification to the model,
thus, incorporates a word-dependent threshold $\delta_i$, in such
a way that the probability of a new occurrence of word $i$ is
proportional to $\max \{n_i, \delta_i\}$. The choice of this set
of thresholds is to a great extent arbitrary. For simplicity we
have chosen an exponential distribution, for which we just have to
specify one parameter, namely the mean $\delta$. The dynamical
effect of this new addition to the model is that words that have
been recently introduced, for which $n_i < \delta_i$, have a
slightly higher competition advantage and their reappearance in
the  text is favored. Since  this parameter  has no influence on
words for which $n_i>\delta_i$, we expect that the power law
region of the distribution, where $n_i \gg \delta_i$, will not be
affected by $\delta_i$.

\begin{figure}
\begin{center}
\includegraphics[width=\columnwidth,keepaspectratio=true]{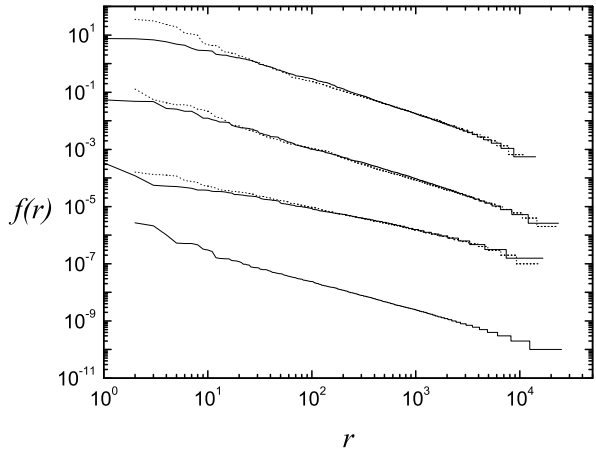}
\caption{As in Fig.~\ref{fig1} the three uppermost curves are
the rank-frequency distributions obtained for David Copperfield,
Don Quijote, and {\AE}neid. The fourth distribution
represents a realization of Simon's model. In this case the
dotted lines are realizations of the model presented in this
paper with the parameters shown in Table \ref{table1}.}
\label{fig2}
\end{center}
\end{figure}

In Fig.~\ref{fig2} we show three realizations of the model that
fit quite well the empirical data already shown in
Fig.~\ref{fig1}. The slope of the distribution in the power-law
region is accurately reproduced for the three literary works.
Moreover, its faster decay in the low-frequency regime is also in
close agreement with the actual linguistic data. Table
\ref{table1} reports the value of the parameters used to fit the
distributions.

We have performed the same analysis using other literary works,
and obtained consistent results. In particular, the model
reproduces Zipf distributions of texts written in languages with
many inflexions, like Latin or Russian for instance, with higher
values of $\nu$ than those required to fit Zipf distributions
obtained from texts written in less inflective languages such as
English or Spanish. In all cases, the faster decay exhibited by
the rank-frequency distributions at high ranks was obtained by
choosing the parameter $\delta$ between $2$ and $4$.

\begin{center}
\begin{table}
\begin{center}
\begin{tabular}{|c|c|c|c|}
\hline Source  & $\alpha_0$              &$\nu$ &$\ \delta \ $
\\      \hline David Copperfield     &
0.03                  & 0.85              & 3          \\
\hline Don Quijote           & 0.05                  &
0.90               & 2          \\      \hline {\AE}neid &
0.25                  & 0.95 & 3          \\      \hline
\end{tabular}
\end{center}
\caption{Model parameters that reproduce the empirical
rank-frequency distributions for the three literary works, used
in the fittings of Fig.~\ref{fig2}.} \label{table1}
\end{table}
\end{center}

We can analytically explain the behavior of our modified Simon's
model starting from Simon's equations for the mean number $P_n(t)$
of words with $n$ occurrences at step $t$. These are \cite{simon}
\begin{equation} \label{P}
P_n(t+1)=P_n(t) +\frac{1-\alpha}{t}[(n-1)P_{n-1}(t)-nP_n(t)]
\end{equation}
for $n>1$, and
\begin{equation} \label{P1}
P_1(t+1)=P_1(t)+\alpha-\frac{1-\alpha}{t} P_1(t).
\end{equation}
Replacing $n$ and $t$ by continuous variables, $n\to y$ and $t\to
t$, and $P_n(t) \to P(y,t)$, Eq.~(\ref{P}) can be approximated by
\begin{equation} \label{Pc}
\partial_t P +\frac{1-\alpha}{t} \partial_y (yP)=0 ,
\end{equation}
while for $P(1,t)\equiv P_1(t)$ we have
\begin{equation} \label{Pc1}
\dot P_1 =\alpha -\frac{1-\alpha}{t} P_1.
\end{equation}

Consider first the situation where the rate at which new words
are added depends on $t$ as $\alpha = \alpha_0 t^{\nu-1}$ ($0<\nu
< 1$), describing a vocabulary that grows less than linearly,
$V\propto t^\nu$. Equations (\ref{Pc}) and (\ref{Pc1}) can be
solved assuming that $\alpha \ll 1$ for all $t$. In such limit, in
fact, the general solution to Eq.~(\ref{Pc}) is given by
\begin{equation} \label{Pya}
P(y,t) = \frac{1}{y} \Pi [yt^{-1}],
\end{equation}
where $\Pi(x)$ is an arbitrary function. For the words that just
appeared once in the text, we have
\begin{equation} \label{P1a}
P_1(t) = At^{-1} + \frac{\alpha_0}{\nu+1} t^\nu,
\end{equation}
with $A$ an arbitrary constant. The second term dominates the
large-$t$ regime. Focusing on this regime and comparing with
Eq.~(\ref{Pya}), we find $\Pi(x) = \alpha_0 x^{-\nu} /(\nu+1)$
and, thus,
\begin{equation} \label{P1b}
P(y,t) =\frac{\alpha_0 }{\nu+1} t^\nu y^{-1-\nu}.
\end{equation}
Taking into account that $P(y,t)$ and $n(r)$ are related
according to $r=\int_n^\infty Pdy$, the Zipf exponent resulting
from Eq.~(\ref{P1b}) is $z=1/\nu>1$. We therefore conclude that
sublinear vocabulary growth, associated with a poorly inflective
language, is able to explain a relatively large value of $z$,
beyond the predictions of the original Simon's model.

The  analytical treatment  of  our second  modification  to the
model requires a  preliminary discussion. In fact, the thresholds
$\delta_i$ cannot be straightforwardly incorporated to Simon's
equations. We can however approximate the dynamical features
associated with them by introducing a few simplifications to the
process. Let us first assume that all the thresholds are set to a
value equal to  their mean, namely $\delta$. Consider now the
events in which a word must be chosen among those written so far.
There are two possibilities: with probability $\gamma$ the word
is chosen among those for which $n_i<\delta$, with uniform
probability, or with probability $1-\gamma$ the word is chosen
among those for which $n_i>\delta$, with probability proportional
to $n_i$. A kind of {\it mean field} approximation of these
dynamical rules would imply that with probability $\gamma$ the
new word is chosen with uniform probability among all the
previous words regardless of $n_i$, and with the complementary
probability the word is chosen with probability proportional to
$n_i$.

In the case of  constant $\alpha$, we introduce this additive
contribution to the evolution of $P_n(t)$ in the following way:
\begin{eqnarray}
P_n(t+1)&=&P_n(t)\nonumber \\
&+&\frac{(1-\alpha)(1-\gamma)}{t}[(n-1)P_{n-1}(t)-nP_n(t)] \nonumber \\
\label{Pad} &+&\frac{(1-\alpha)\gamma}{t}[P_{n-1}(t)-P_n(t)].
\end{eqnarray}
The evolution of $P_1$ is still given by Eq.~(\ref{P1}).
Particular solutions to the continuous version of Eq.~(\ref{Pad}),
indexed by the parameter $\lambda$, read
\begin{equation} \label{lambda}
P(y,t) = a_\lambda t^\lambda
[(1-\gamma)y+\gamma]^{-1-\lambda/(1-\alpha)(1-\gamma)},
\end{equation}
where $a_\lambda$ is an arbitrary constant.

In the large-$t$ regime and within the assumption $\alpha\ll 1$ we
can put together the effect of $t$-dependent $\alpha$, $\alpha(t)=
\alpha_0 t^{\nu-1}$, and of the additive process. In fact, the
large-$t$ dominant contribution in Eq.~(\ref{P1a}) is compatible
with solution (\ref{lambda}) for $\lambda=\nu$. We obtain
\begin{equation}
P(y,t) = \frac{\alpha_0}{1+\nu} t^\nu
[(1-\gamma)y+\gamma]^{-1-\nu/(1-\gamma)}.
\end{equation}
Zipf's law results
\begin{equation}
n = \frac{1}{1-\gamma} [(r/r_0)^{-(1-\gamma)/\nu} -\gamma] ,
\end{equation}
with $r_0(t)= \alpha_0 t^\nu/(1+\nu)$. The distribution shows a
cutoff at $r=gr_0$, with $g=\gamma^{-\nu/(1-\gamma)}$, which
explains its faster decay for large ranks.

In summary, we have investigated the origin of Zipf's law, which
stands for the most basic statistical pattern found in written
human language. Our results confirm that the rank-frequency
distribution of words is mainly the consequence of multiplicative
processes underlying the language generation process. Moreover,
we have been able to associate the finer details of empirical
distributions with two key subprocesses that participate in text
production. First, a sublinear vocabulary growth law that
integrates the effect of the inflective structure of language.
Second, context-dependent dynamics and activation thresholds
related to the dynamical effect of newly introduced words. Our
main conclusion is that linguistically sensible modifications of
Simon's model eliminate the slight deviations between the results
of the original model and actual Zipf distributions. This gives
additional support to the interpretation of Zipf's law as a
linguistically significant feature of written texts.

\end{document}